# Quantifying quasiparticle chirality in a chiral topological semimetal

Jiaju Wang[1], Jaime Sánchez-Barriga[2,3], Amit Kumar[1], Markel Pardo-Almanza[1], Jorge Cardenas-Gamboa[4], Iñigo Robredo[5], Chandra Shekhar[6], Daiyu Geng[1], Emily C. McFarlane[1], Martin Trautmann[1,7], Enrico Della Valle[8], Moritz Hoesch[9], Meng-Jie Huang[9], Jens Buck[9], Vladimir N. Strocov[8], Annika Johansson[1,7,13], Stuart S. P. Parkin[1], Claudia Felser[6], Maia G. Vergniory[10,11,12] & Niels B. M. Schröter[1,7,13,*]

[1] Max Planck Institute of Microstructure Physics, Weinberg 2, 06120 Halle (Saale), Germany

[2] Helmholtz-Zentrum Berlin für Materialien und Energie, Elektronenspeicherring BESSY II, Albert-Einstein-Strasse 15, 12489 Berlin, Germany

[3] IMDEA Nanoscience, C/ Faraday 9, Campus de Cantoblanco, 28049 Madrid, Spain

[4] Leibniz Institute for Solid State and Materials Research, IFW Dresden, Helmholtzstraße 20, 01069 Dresden, Germany

[5] Smart Materials Unit, Luxembourg Institute of Science and Technology (LIST), Avenue des Hauts-Fourneaux 5, L-4362 Esch/Alzette, Luxembourg

[6] Max Planck Institute for Chemical Physics of Solids, 01187 Dresden, Germany

[7] Institute of Physics, Martin Luther University Halle-Wittenberg, 06120 Halle (Saale), Germany

[8] Swiss Light Source, Paul Scherrer Institut, Villigen, CH-5232 PSI, Switzerland

[9] Deutsches Elektronen-Synchrotron DESY, 22607 Hamburg, Germany

[10] Donostia International Physics Center, Manuel Lardizabal Ibilbidea 4, 20018 Donostia-San Sebastian, Gipuzkoa, Spain

[11] Département de Physique et Institut Quantique, Université de Sherbrooke, J1K 2R1 Sherbrooke, Québec, Canada

[12] Regroupement Québécois sur les Matériaux de Pointe (RQMP), H3T 3J7 Québec, Canada

[13] Halle-Berlin-Regensburg Cluster of Excellence CCE, 06120, Halle (Saale), Germany

*Corresponding author. Email address: niels.schroeter@mpi-halle.mpg.de

# Abstract

Recently, the projection of the electron's spin on its crystal momentum has been proposed as a metric to quantify electronic chirality of Bloch states in crystals, which is expected to affect a wide range of physical properties, such as magnetoelectric and optical responses. However, a direct experimental quantification of this chirality metric over an entire iso-energy surface has remained elusive. Here, we have used spin- and angle-resolved photoemission spectroscopy to directly probe the electronic chirality by measuring the bulk spin texture of Kramers-Weyl and Weyl cones in RhSi, a chiral topological semimetal with strong spin-orbit coupling (SOC). After quantifying the SOC splitting of Weyl cones, we determine their spin direction along different azimuthal angles to extract energy dependent the deviations (up to ~40°) from perfect parallel spin-momentum locking. From these deviations we define an energy-dependent normalized electron chirality density (NECD), a directly accessible metric of bulk electronic chirality. In RhSi, the NECD decreases from 1 at the Kramers-Weyl point to ~0.8 at ~200 meV below it. Finally, we show that this experimentally grounded NECD provides predictive power for magneto-optical and transport responses of chiral materials, exemplified by the longitudinal Edelstein effect.

# Introduction

Chirality, defined as the breaking of all improper symmetry operations, enables unconventional electronic[1–12], optical[13–15], and magnetic[16–18] phenomena in solids. Yet, this binary definition of chirality provides little guidance for predicting the magnitude of unconventional phenomena expected to appear in chiral materials. To tackle this problem, recent theoretical studies have moved towards a modern theory of chiralization[19], proposing several order parameters and metrics to quantify chirality on a continuous scale, including spinful quantities such as electron chirality[20,21], electric toroidal monopoles[22,23], and spinless electric toroidal quadrupoles[24]. Among these metrics, electron chirality, defined as proportional to $k \cdot \sigma$ for linear bands in the non-relativistic limit[20], stands out as a directly measurable quantity defined for individual quasiparticle states. Previous research found that electron chirality directly influences the intrinsic energy gap between enantiomers[21,25,26]. Furthermore, it has been suggested to affect optical properties and magnetoelectric transport such as current-induced optical activity[24,27,28] and the spin Edelstein effect[24,29–32]. However, despite these theoretical connections, electron chirality has never been experimentally quantified in chiral materials, and there has been no explicit theoretical calculation quantitatively linking the degree of electron chirality and the magnitude of transport responses. Consequently, whether electron chirality can serve as a predictive parameter remains experimentally and theoretically unresolved.

To enable experimental quantification of electron chirality of quasiparticles in metals, we define a normalized electron chirality density (NECD), which is directly applicable to electronic states in crystals:

$$\text{NECD} = \widehat{k} \cdot \sigma \qquad (1)$$

Where $\widehat{k} = k/|k|$ is the crystal momentum unit vector, and $\sigma$ represents the Pauli matrices acting on physical spin. Figure 1a shows the NECD of several spin-textures described by prototypical $k \cdot p$ Hamiltonians in the small-$k$ limit in the vicinity of a high-symmetry point (Fig. 1a). The electronic states on the Au (111) surface near $\bar{\varGamma}$ exhibit perpendicular Rashba spin-momentum locking (SML), which have NECD = 0. In trigonal tellurium (Te), for the Weyl node at *A* point in the Brillouin zone, the spin is parallel to momentum only along a single screw axis[8,10], yielding an NECD less than or equal to 1. For cubic B20 crystals such as RhSi, due to the threefold rotation axes along ⟨111⟩ and twofold screw axes along ⟨100⟩, spins in all directions are locked perfectly parallel to momentum for Weyl fermions at the $\varGamma$ point [8,11,33], resulting in an NECD of 1 (the largest possible value) over the entire iso-energy surface enclosing the Weyl fermion. Nevertheless, further away from $\varGamma$, higher order terms in the Hamiltonian lead to deviations from parallel SML (Fig. 1b), and the NECD decreases. It is not yet experimentally known how strong this decrease would be and what effect it could have on transport responses. This further motivated us to quantify the NECD experimentally in a prototypical chiral topological semimetal crystallizing in the B20 structure, RhSi, using spin- and angle-resolved photoemission spectroscopy (spin-ARPES).

Compared to other well-studied transition metal silicides such as CoSi, RhSi has stronger spin-orbit coupling (SOC), leading to larger band splittings, and is therefore expected to result in stronger spin-dependent magnetoelectric responses. Another advantage for RhSi over other B20 compounds is that the Weyl bands remain relatively linear over a wide energy range, and their dispersions are separated from other bands near $\varGamma$ (left panel of Fig. 1b). This allows us to directly probe Weyl cone spin textures along multiple *k* directions with spin-ARPES. Whereas in other B20 compounds such as PtGa[9,34], the spin texture of Fermi arc surface states has been used as an indirect probe of the unconventional multifold fermion spin textures[9], and measurement of the Weyl cone spin texture has only been achieved along one high-symmetry direction. Despite these previous experimental characterizations of spin textures in B20 compounds[9,34], a direct measurement of the complete spin texture of the bulk Weyl cones across an entire iso-energy surface has remained challenging. Consequently, deviations from parallel SML, which are directly related to the NECD, have not yet been experimentally quantified. From the above analysis, we conclude that RhSi is an ideal candidate for probing the spin texture of Weyl cones and studying electron chirality due to its strong SOC and linear Weyl cone dispersion.

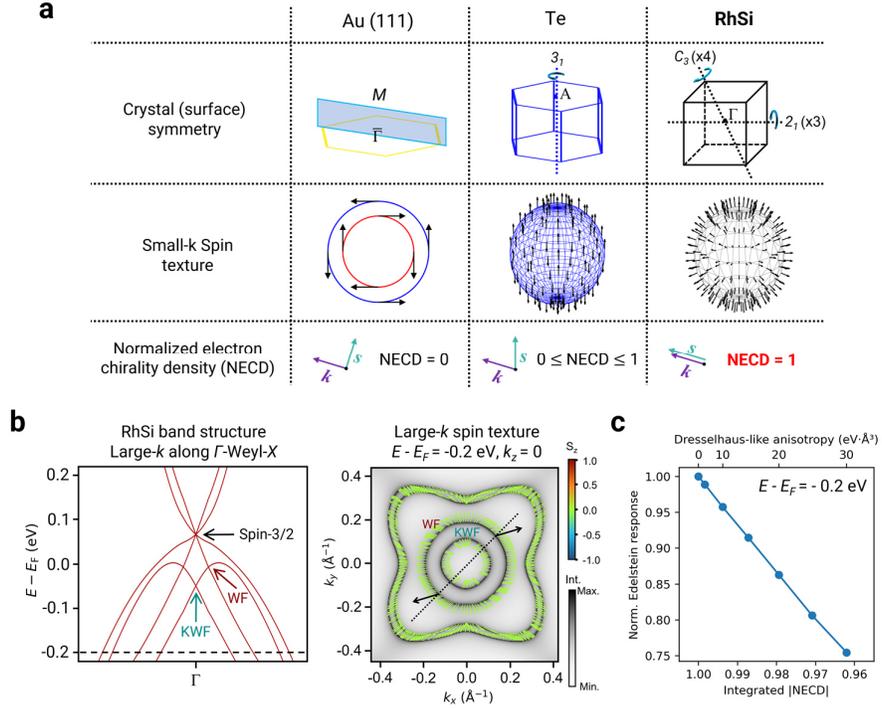

**Fig. 1: Spin-momentum locking and electron chirality density. a**, A summary of different crystal symmetries in selected compounds and corresponding Brillouin zones, the resulting spin textures in the small-$k$ limit, and NECD values. The B20 compound RhSi exhibits parallel SML, yielding an NECD of 1 for all small $k$ values. **b,** DFT calculated band structure and iso-energy-contour with spin texture for large $k$ values. Left panel shows the band structure along the $k$-path $\Gamma$-type I Weyl point-$X$. The crossings are a spin-3/2 multifold fermion and a Kramers-Weyl fermion (labelled as KWF) at $\Gamma$, and a type I Weyl fermion (labelled as WF) close to $\Gamma$. Dashed line indicates the energy $E - E_F = -0.2$ eV of the contour plot in the right panel, which also shows spin direction. Momentum-dependent spin deviations from parallel SML are present in the two inner circular bands corresponding to the Kramers-Weyl and Weyl cones. **c,** Linear correlation between Edelstein response and electron chirality at $E - E_F$ = -0.2 eV, calculated using a $\bm{k} \cdot \bm{p}$ model with a Dresselhaus-like anisotropy term.

In this work, spin-ARPES was used to experimentally quantify electron chirality of Weyl quasiparticles in RhSi. First, the spectral features of angle-resolved photoemission spectroscopy (ARPES) corresponding to the Kramers-Weyl fermion and Weyl fermion were identified, and the SOC-splittings of Weyl cones along the $\Gamma$-$X$ direction were measured. Then, the $k$-dependent spin deviation from parallel SML was characterized along different azimuthal angles, and compared with density functional theory (DFT) calculations. Finally, using our definition of NECD and the spin deviation angles, we directly quantify Weyl quasiparticle chirality in chiral crystals for the first time. Combined with our $\boldsymbol{k} \cdot \boldsymbol{p}$ calculations, which indicate the decrease of the Edelstein effect response with decreasing NECD, we establish NECD as a measurable parameter linking chirality and magnetoelectric transport. Beyond the specific case of RhSi, the broader significance of this work lies in reframing chirality for quantum materials as a quantifiable and experimentally benchmarkable property, rather than only a symmetry label. Conceptually, our study establishes a route to extract chirality directly from spin texture, which can renormalize chirality-dependent magnetoelectric responses. More broadly, this provides a practical framework for connecting microscopic quasiparticle structure to macroscopic functionality, which is relevant for the design and comparison of chiral materials for magnetoelectric and spintronic applications.

## Results

**Theoretical calculations of spin deviations and the Edelstein response in RhSi**

The DFT calculations shown in Figure 1b provide important insights on the position of Weyl points in RhSi and the magnitude of spin deviation from parallel SML that can guide our experiments. In the band structure (left panel of Fig. 1b, see Figure S1 in Supplementary Information (SI) for more details), there is a Kramers-Weyl point at $\Gamma$ and a type I Weyl point close to $\Gamma$ [35], both separated from the four-fold Rarita-Schwinger-Weyl (spin-3/2) fermion. Because the Weyl points sit below the Fermi energy ($E_F$), and the band structure near the Fermi-level is relatively complex, the only cleanly accessible parts of the band structure in spin-ARPES experiments are the Weyl cones below these points. Here, we choose a representative binding energy of 0.2 eV, at which the contour plot is shown in the right panel of Figure 1b ($k_z$ = 0). The inner circular band corresponds to the Kramers-Weyl cone, and the outer circular band is the Weyl cone. It can be seen that the spin deviations are momentum dependent, meaning that it is necessary to probe the spin along several different $k$ directions to fully understand the spin texture and its electron chirality. It should be noticed that the SOC-splitting between the Kramers-Weyl cone and Weyl cone is larger than other bands below $E_F$, which is advantageous for spin-ARPES measurements.

The spin deviations shown above can directly affect physical properties such as the Edelstein effect in RhSi. Our $k \cdot p$ simulations show that the magnitude of the longitudinal Edelstein response (longitudinal Edelstein spin density) from the Weyl-cones (not considering other bands in the Brillouin zone) at binding energy 0.2 eV decreases with decreasing integrated NECD (integrated over the full iso-energy surfaces enclosed by the Weyl point and Kramers-Weyl point, Fig. 1c), with an approximately linear correlation. We model the decrease in NECD by including a higher-order-in-$k$ Dresselhaus-like anisotropy term in the Hamiltonian (see Figure S2 in SI). Although this model does not simulate the complicated band contour at $E_F$ (see Figure S3 in SI), and does not incorporate the full complexity of realistic band structures or scattering mechanisms relevant to experiments, it demonstrates the general trend that decreasing NECD reduces the Edelstein response, and is instructive to understand how NECD is quantitatively linked to the magnitude of magnetoelectric transport.

**Large SOC splitting between Kramers-Weyl and Weyl cones**

Following what we learned from theoretical calculations, we first present experimental evidence showing the large SOC-splitting of the Weyl cones in RhSi. Figure 2a shows a Fermi surface (FS) measured with bulk-sensitive soft X-ray ARPES at $hv$ = 550 eV, probing the bulk bands of RhSi near the $\Gamma$ point. To confirm the assignment of the correct momentum along the out-of-plane direction $k_z$, additional soft X-ray ARPES data measured at different photon energies over multiple Brillouin zones are shown in Figure S4 of the SI. The shape of the FS in both Figure 2a and Figure S4 resembles that of DFT calculations in Figure 1b, which further confirms the correct identification of bulk bands near the $\Gamma$ point in our experiment.

Since spin-resolved ARPES at soft-X-ray energies is challenging due to low photoemission cross-sections, in this work we performed spin-ARPES experiments at lower photon energies of $hv = 40$ eV. Importantly, the resulting electronic band dispersion closely matches the results from soft-X-ray ARPES as can be seen in Figure 2b, which indicates that these measurements correspond to equivalent final state momenta near $\Gamma$. In the experiments we find two well separated bands. The comparison of vacuum ultraviolet (VUV) and soft X-ray data in Figure 2b shows that the positions of the two bands in VUV agrees well with those of the bulk bands measured with soft X-rays. Comparison with DFT calculations shows that the band at lower energy corresponds to the Kramers-Weyl cone and the Weyl cone, thus referred to as KW+W, while the band at higher energy corresponds to the middle bands (MB in right panel of Fig. 2b) in the band structure. The Kramers-Weyl cone has lower energy than the Weyl cone, as indicated in the DFT panel in Figure 2b. A spin-resolved energy distribution curve (EDC) at a momentum position indicated by a dashed green circle in Figure 2a (which is around 0.23 Å$^{-1}$ away from $\Gamma$ along the negative $\Gamma$-$X$ direction) is shown in Figure 2c. The spin-resolved EDC cuts through the two VUV bands.

Although the SOC splittings of the Kramers-Weyl cone and the Weyl cone are not observable in the spin-integrated FS (Fig. 2a) or cuts (Fig. 2b), the higher resolving power of spin-ARPES can reveal these splittings and help us separate the spin of the Kramers-Weyl and Weyl cones. Figure 2c shows the in-plane $x$ component of spin, whose direction is along $\Gamma$-$X$. The upper panel displays the spin-up and spin-down intensity spectra ($I_+$ and $I_-$), and the $x$ component spin polarization is shown in the lower panel. Here the spin up and down intensities can be extracted from the spin polarization and the total intensity (see Methods section), and represent the number of photoelectrons which have spin up and down eigenstates. Conversely, the asymmetry of spin up and down intensities is equal to spin polarization (see Methods section). The spin up (down) direction for the $x$ component is defined as parallel (antiparallel) to the positive $k_x$ direction. The spin up and spin down data points correspond to red and blue triangles, respectively. The red and blue lines represent a corresponding fit to those data. Similarly, in the spin polarization, dots represent the data points and the line refers to the polarization obtained from fitted intensities. The fitting process is described in the Methods section.

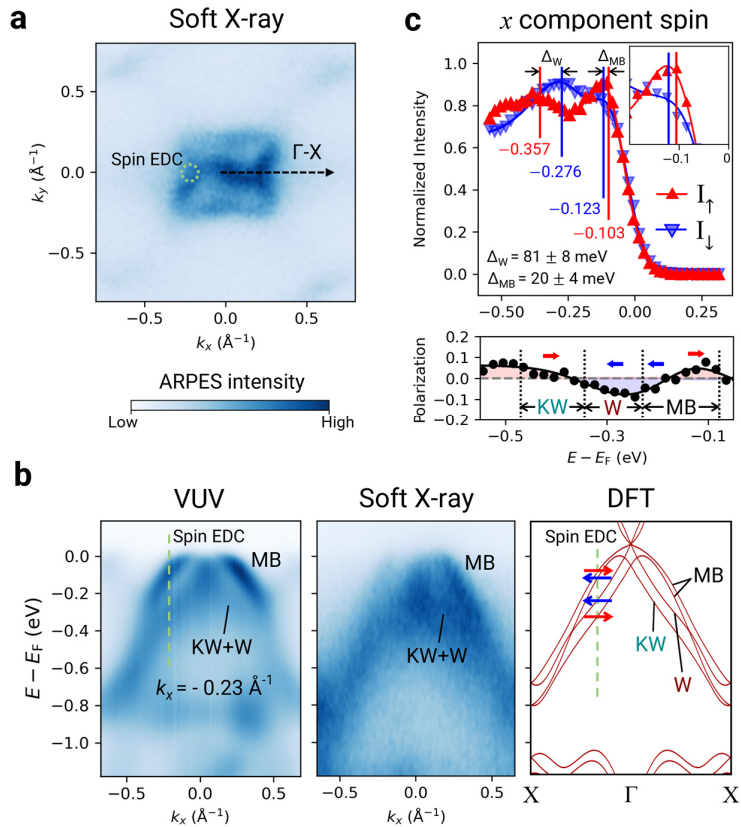

**Fig. 2: Large SOC-splitting of bulk bands in RhSi. a**, Soft X-ray FS of RhSi measured with $hv = 550$ eV, showing a similar shape to the calculated constant-energy contour in Figure 1b. The spin-resolved EDC measured at 40 eV is indicated by the dotted green circle, which is along the $\Gamma$-$X$ direction. **b**, Left panel is the cut along $\Gamma$-$X$ at 40 eV, showing that this EDC (dashed green line) cuts through two bands. Comparison with soft X-ray cut (middle panel) and DFT calculation (right panel) show that the bands correspond to the Kramers-Weyl cone and the Weyl cone (KW+W), and the middle bands (MB). **c**, In-plane $x$ component spin intensities and polarization for the spin-resolved EDC, measured at 40 eV. Inset in the upper panel shows a zoomed-in view of the MB region. Spin directions are labelled on the polarization, which are in the same direction as those calculated with DFT in the right panel of **b**.

Focusing on the spin intensities, there is a clear shift between the peaks corresponding to spin up and spin down around $E - E_F$ = -0.3 eV, indicating the expected spin-splitting of Kramers-Weyl and Weyl cones. The spin-splitting between the middle bands is smaller, which is enlarged in the inset of the upper panel in Figure 2c for clarity. According to fitting results, the splitting of KW+W and MB are calculated as 81 ± 8 meV and 20 ± 4 meV, respectively. According to the DFT calculations, the SOC-splittings at $k_x$ = 0.2 Å$^{-1}$ are 83 meV and 23 meV, which are both consistent with experimental values within their errors. It should be noted that for the *x* component, Kramers-Weyl and Weyl cones have spin up and spin down, respectively, while for the MB, the band with lower energy is spin down, and the band with higher energy is spin up. These spin directions are illustrated with red and blue arrows in the spin polarization (bottom panel of Fig. 2c). Comparing the spin directions with the spin texture calculations (labeled in the right panel of Fig. 2b), we find that both show an 'up down down up' pattern from low to high energy. The *y* and *z* components of spin are displayed in Figure S5 of the SI, which have smaller spin-splittings.

Spin-resolved measurements testing the time reversal symmetry of the Weyl cones are shown in Figure S6 of the SI. These measurements show a flip in the spin direction for EDCs at opposite *k* points, consistently showing that the measured spin polarization is due to SOC, rather than final state effects or residual magnetism.

**Spin-deviations of Weyl quasiparticles along different *k* directions**

Having identified the Weyl cones and their spin splitting, we can now extract their spin texture along several *k* directions between *Γ-X* and *Γ-Y* by rotating the azimuthal angle. Due to the matrix element effect[36,37], the photon energy 20 eV was used instead of 40 eV to increase the intensity of the KW+W band (see Figure S7). The similar band dispersions (Figure S7) indicate that 20 eV and 40 eV have similar $k_z$ values. Taking *Γ-X* as 0° and *Γ-Y* as 90°, spin measurements were carried out at azimuthal angles of 90°, 67°, 45°, and 22°.

Figure 3 displays the spin-resolved EDC positions and the spin data. The out-of-plane spin component show near-zero spin polarization (Figure S8 in the SI), thus here we focus on the in-plane spin component. Figure 3a shows the spin-integrated cuts through the $\Gamma$ point at different azimuthal angles, where 0° indicates the $\Gamma$-X and 90° the $\Gamma$-Y direction. The corresponding k-positions of the spin-resolved EDCs are indicated by a black dashed line. At 90°, the EDC cuts through both a surface state (labelled with S) around $E - E_F = -0.1$ eV and the KW+W band around $E - E_F = -0.3$ eV, as shown in Figure 3a. The surface state appears as a shoulder in the EDC and is separated from the KW+W. The $k_z$ scan of the cut at 90° in Figure S9 provides evidence for the bulk nature of KW+W band and the surface nature of the surface state. For EDCs extracted from the azimuthal cuts from 67° to 22°, the KW+W band is the only visible peak in the EDCs.

Figure 3b-d shows the spin-resolved measurements from which the spin deviation from parallel SML can be deduced. Figure 3b shows the in-plane spin component parallel to momentum ($S_{//}$), where spin up (down) is defined as parallel (antiparallel) to the positive $k_{//}$ direction. According to results in Figure 2, the spin asymmetry arises from the Kramers-Weyl cone at lower energy and the Weyl cone at higher energy (labelled in 90° $S_{//}$ panel in Fig. 3b). Importantly, from 90° to 22°, it can be seen that the asymmetry of $S_{//}$, which is directly proportional to the spin polarization, decreases. Figure 3c shows the in-plane spin component perpendicular to momentum ($S_\perp$). Spin up (down) is defined as parallel (antiparallel) to the positive $k_\perp$ direction. Contrary to $S_{//}$, from 90° to 22°, the spin asymmetry of $S_\perp$ increases. Figure 3d shows spin polarizations of the parallel ($P_{//}$) and perpendicular ($P_\perp$) components. According to fitting results, the regions corresponding to Kramers-Weyl and Weyl cones are labelled in the spin polarizations, and green and purple arrows represent the approximate magnitude of $P_{//}$ and $P_\perp$, respectively. It can be seen that the ratio $P_\perp / P_{//}$ increases from 90° to 22°. For parallel SML, we would expect $P_\perp = 0$, and the larger the spin deviation from parallel SML, the larger the ratio $P_\perp / P_{//}$. Therefore, qualitatively, this increase in $P_\perp / P_{//}$ indicates increasing spin deviation from 90° to 22° (decreasing spin deviation with increasing azimuthal angle).

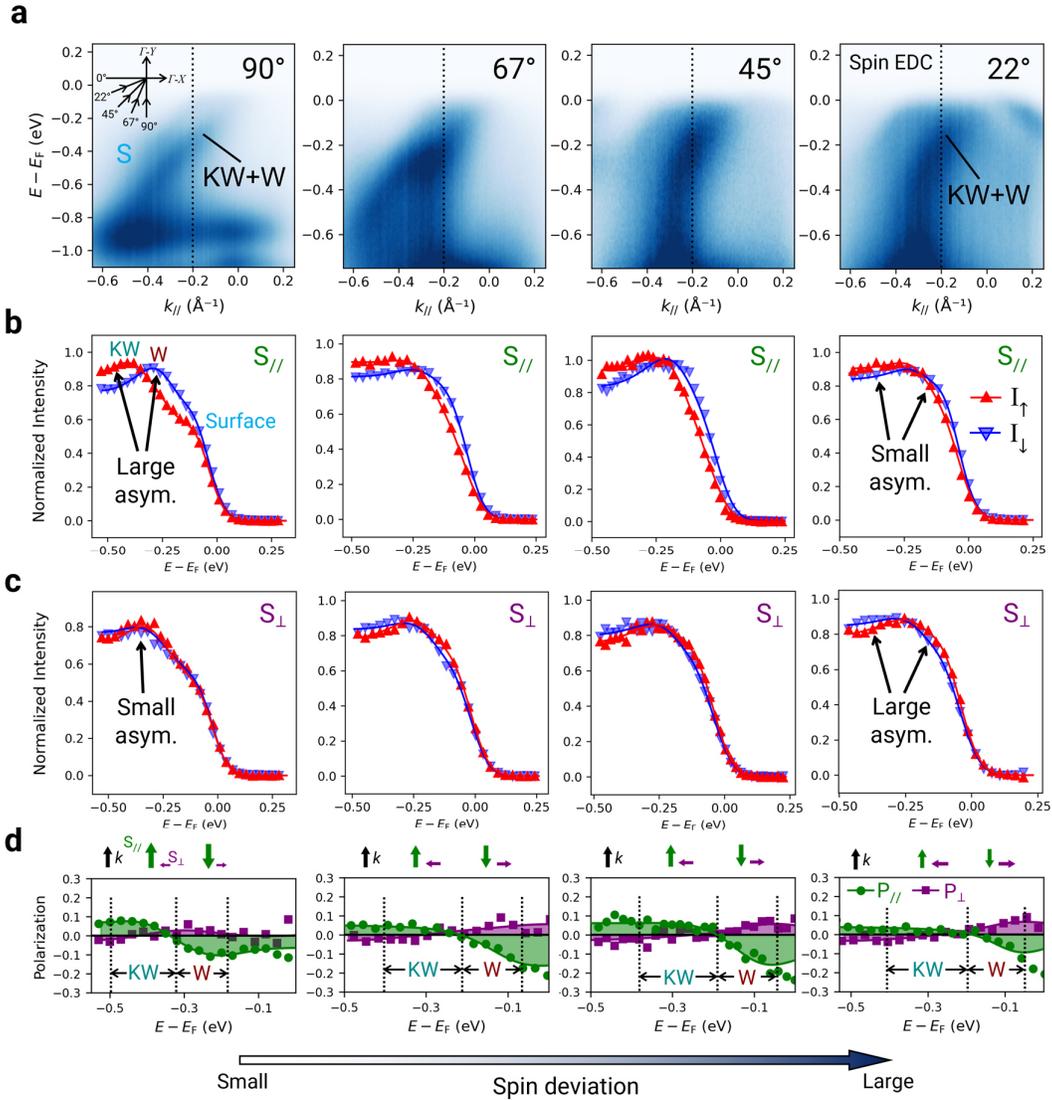

**Fig. 3: Spin-deviations at different azimuthal angles. a,** VUV-ARPES cuts measured using 20 eV at azimuthal angles of 90°, 67°, 45° and 22°. The zero point in each cut corresponds to $\Gamma$. Spin-resolved EDCs are labelled by black dotted lines. For the EDC at 90°, a surface state is present near $E_F$ indicated by the blue S, separated from the Weyl fermions. **b,** Parallel-to-momentum component of in-plane spin up and down intensities at azimuth angles corresponding to those in **a**. The asymmetry decreases from 90° to 22°. **c,** Perpendicular-to-momentum component of in-plane spin up and down intensities. The asymmetry increases from 90° to 22°. **d,** In-plane spin polarizations parallel and perpendicular to momentum. Arrows above the plots illustrate the $k$ direction and the relative magnitudes of the spin parallel and perpendicular to momentum. The ratio $P_\perp/P_{//}$ decreases from 90° to 22°, indicating an increasing spin deviation.

To test for final state effects, spin measurements from 67° to 22° were repeated using photon energy 40 eV (see Figure S10). The results are comparable with those in Figure 3. This indicates that no final state effects are present when changing the photon energy from 40 to 20 eV, suggesting that we are measuring the initial state spin polarization[38,39].

**Quantifying quasiparticle chirality in RhSi**

To quantify quasiparticle chirality in RhSi, we extracted the value of the spin angle from the data shown in Figure 3 (see Methods section for details about spin angle extraction and error estimation). Figure 4a indicates the momentum positions where the spin-direction was measured.

The extracted experimental spin texture with errors and comparison with DFT spin texture is shown in Figure 4b. The Weyl cone was consistently probed around $E - E_F$ = -0.2 eV (top panels), and the Kramers-Weyl cone was probed at $E - E_F$ = -0.3 eV (bottom panels). The extracted spin directions are represented by dark red arrows for the Weyl cone and dark cyan arrows for the Kramers-Weyl cone. The errors are labelled by dotted lines with the same color as the arrows. It can be seen from the experimental data that a consistent clockwise spin-deviation from ideal parallel SML can be detected for both Kramers-Weyl and Weyl cones at 67°, 45°, and 22°, and this deviation is larger than the estimate of our measurement uncertainty. The DFT-calculated spin angles at these azimuthal angles are highlighted in the zoomed-in constant energy contours in the right panels of Figure 4b, which also exhibit clockwise spin deviations, consistent with experimental results. This clockwise deviation is related to the sign of higher-order terms in the Hamiltonian. As demonstrated in Figure S11 in the SI, when the sign of the Dresselhaus-like term is flipped, the spin deviations change from clockwise to anticlockwise. The spin deviation angle is calculated as the angle between the actual spin state and a spin state with perfect parallel SML, and the absolute value of NECD, |NECD|, is equal to cos(spin deviation angle). The above parameters for Kramers-Weyl and Weyl cones obtained

from experiment and DFT calculations are compared in Figure 4c, and their values are displayed in Table 1. In both experiment and DFT calculations, the spin deviations for Kramers Weyl and Weyl cones have similar values, due to the fact that we are measuring them at different energy values. The main difference is that in the experimental data, the spin deviation angle decreases with increasing azimuthal angle, while it increases in DFT calculations. From the above analysis, we can see that the DFT calculation can successfully predict the spin deviation direction and its order of magnitude. However, it cannot reproduce the specific experimental spin texture and spin deviations accurately.

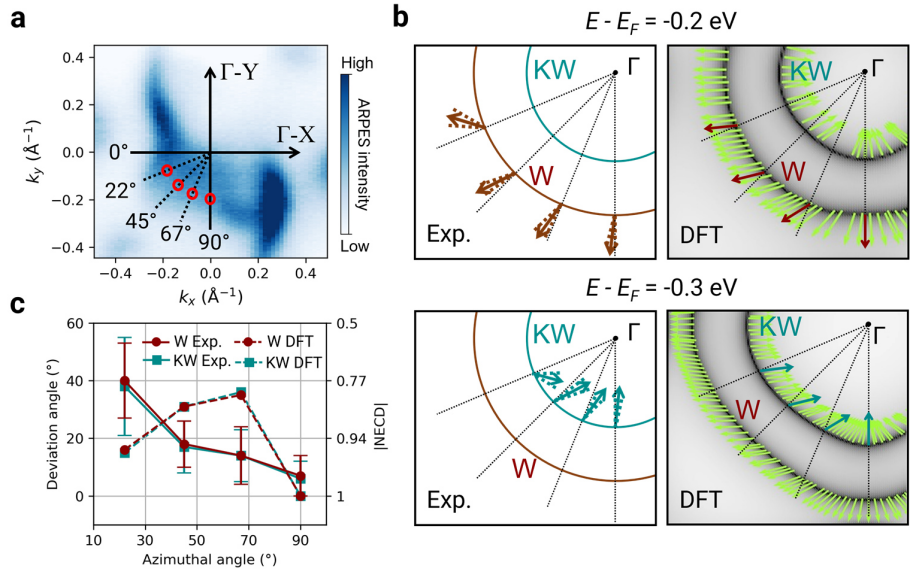

**Fig. 4: Quantification of Weyl quasiparticle chirality in RhSi. a**, VUV contour at $E - E_F = -0.2$ eV of RhSi measured at 20 eV. Red circles indicate spin-resolved EDC positions in Figure 3. **b,** Left panels show the spin textures extracted from experiment of the Weyl cone and Kramers-Weyl cone at $E - E_F$ values -0.2 eV and -0.3 eV, respectively. Dashed lines indicate the error of each spin angle. Right panels show the same spin textures calculated with DFT. All spin textures show a clockwise spin deviation. **c,** Line plot showing how the spin deviation angle and |NECD| changes with azimuthal angle according to experimental and theoretical results. Experimentally, with increasing azimuth angle, the spin deviation angle decreases, hence the |NECD| increases.

In summary, the measurements show a clockwise deviation of spin in the Weyl cones for azimuthal angles between $\Gamma$-$X$ and $\Gamma$-$Y$, which matches the deviation direction in DFT calculations. Experimentally, the deviation is the largest at an azimuth angle of 22° and decreases at larger azimuthal angles. Correspondingly, the |NECD| increases from around 0.8 to almost 1 for azimuthal angles going from 22° to 90°.

**Table 1.** Summary of experimental and theoretical spin deviation angles of the Kramers-Weyl (KW) and Weyl (W) cones, and the |NECD| values at different azimuthal angles. Experimental values are shown with errors.

| Azimuthal angle (°) | Method | KW cone spin deviation (°) | $|NECD|_{KW}$ | W cone spin deviation (°) | $|NECD|_W$ |
|---|---|---|---|---|---|
| 90 | Experiment | 6 ± 6 | 0.99 ± 0.01 | 7 ± 7 | 0.99 ± 0.01 |
|    | DFT        | 0     | 1.00        | 0     | 1.00        |
| 67 | Experiment | 14 ± 9 | 0.97 ± 0.04 | 14 ± 10 | 0.97 ± 0.04 |
|    | DFT        | 36    | 0.81        | 35    | 0.82        |
| 45 | Experiment | 17 ± 9 | 0.96 ± 0.05 | 18 ± 8 | 0.95 ± 0.04 |
|    | DFT        | 31    | 0.86        | 31    | 0.86        |
| 22 | Experiment | 38 ± 17 | 0.79 ± 0.18 | 40 ± 13 | 0.77 ± 0.15 |
|    | DFT        | 15    | 0.97        | 16    | 0.96        |

## Discussion

In summary, a comprehensive study of the RhSi spin texture is carried out in this work. Starting from the 40 eV spin-resolved EDC taken from a cut along $\Gamma$-$X$, the VUV band corresponding to Kramers-Weyl and Weyl cones has been identified. By conducting measurements of the spin texture of both Weyl cones at various azimuthal angles, we have showed that the spin deviation from parallel SML decreases with increasing azimuthal angle between $\Gamma$-$X$ and $\Gamma$-$Y$. Quantification of the spin deviation angle shows that it can reach up to ~40°, and can cause the |NECD| to drop from 1 to around 0.8.

Our results have established NECD as a parameter that correlates microscopic spin textures with macroscopic magnetoelectric transport. For example, in Edelstein effect transport experiments, by comparing similar B20 compounds such as RhSi and CoSi, one can tune the $E_F$ position and SOC strength, thereby modifying the NECD, which affects the magnitude of the Edelstein response. From this example, we can see that quantification of electron chirality provides a practical route to systematically control the Edelstein effect and potentially other transport properties that depend on Fermi-surface spin-textures, such as the Josephson diode effect[40]. More broadly, our work suggests that electron chirality can serve as an experimentally observable parameter for tuning magnetoelectric transport in chiral topological semimetals through designing their band structure and spin texture. It should be noted that for direct magnetoelectric transport measurements, homochiral thin films of chiral topological semimetals are required to achieve high current densities, which is still under pursuit.

Beyond spin, similar deviations from parallel orbital-momentum locking are anticipated in the orbital texture of RhSi, which would have similar influences on the orbital Edelstein effect[29], and can also indicate unconventional Berry curvature and quantum metric distributions, potentially affecting properties such as the nonlinear planar Hall effect[41,42] and the circular photo-galvanic effect[13,43,44]. In brief, spin and orbital texture deviations from idealized theoretical models, such as perfect-Weyl spin-momentum locking, are important factors that can affect the physical responses of chiral materials and can be used to quantify the chirality of quasiparticles.

## Methods

### RhSi crystal growth and surface preparation

Single crystals of RhSi were grown from the melt using the vertical Bridgman crystal growth technique. First, a polycrystalline ingot of the desired composition (RhSi) was prepared by pre-melting highly-pure stoichiometric amounts of respective metals under argon atmosphere using an arc melt technique. The crushed powder was then placed inside a custom-designed sharp-edged alumina tube, which was again sealed inside a tantalum tube in argon atmosphere. First, the sample was heated to 1,550 °C at a rate of 200 °C h$^{-1}$ and held there for 10 h. Then the ampule was slowly pulled to the cold zone, down to 1,100 °C at a rate of 0.8 mm h$^{-1}$. The temperature was controlled by attaching a thermocouple at the bottom of the Bridgman ampule. A single crystal with average dimensions of 9 mm in length and 6 mm in diameter was obtained. Before ARPES measurements, the RhSi single crystal surface was prepared *in-situ* in an ultrahigh vacuum chamber with a base pressure better than $2 \times 10^{-9}$ mbar. The surface preparation process consists of repeated cycles of sputtering (Ar$^+$, 1 keV, ~$1 \times 10^{-5}$ mbar) and annealing (970 K). A sharp LEED pattern was obtained after preparation, indicating that the surface conditions are suitable for photoemission measurements.

**Spin and angle resolved photoemission spectroscopy**

Spin-ARPES and ARPES measurements were carried out at the U125-PGM beamline of the synchrotron radiation source BESSY-II in Helmholtz Zentrum Berlin. Samples were measured at room temperature with p-polarized light. Photoelectrons were detected with a Scienta R4000 analyzer at the spin-ARPES end station, and the base pressure of the experimental setup was better than $1 \times 10^{-10}$ mbar. The angular and energy resolutions were set to be 0.1° and 5 meV, respectively. For spin-resolved measurements, a Rice University Mott-type spin polarimeter was used, operated at 25 kV and capable of detecting all three components of the spin polarization. The energy and angular resolutions of spin-ARPES measurements were 45 meV and 0.75°, respectively.

**Soft X-ray angle resolved photoemission spectroscopy**

Soft X-ray ARPES experiments were performed at the SX-ARPES end station[45] and the ASPHERE III end station at the Variable Polarization XUV Beamline P04 of the PETRA III storage ring at DESY (Hamburg, Germany). The sample was held at around 20 K under a pressure lower than $2 \times 10^{-10}$ mbar. The angular resolution was about 0.1°, and the combined analyzer and beamline energy resolution ranged from 60 to 180 meV for photon energies between 360 eV and 1.021 keV.

**Analysis, fitting, and error calculation of spin-resolved spectra**

From the raw counts of the spin detectors, $I_+$ and $I_-$, the spin polarization was calculated using:

$$P = \frac{1}{S} \frac{I_+ - I_-}{I_+ + I_-} \quad (2)$$

Where $S$ is the Sherman function with a value of 0.12. To obtain the spin up and spin down intensities, $I_\uparrow$ and $I_\downarrow$, the following equations were used:

$$I_\uparrow = \frac{1}{2}(I_{tot} + P * I_{tot}), \; I_\downarrow = \frac{1}{2}(I_{tot} - P * I_{tot}) \quad (3)$$

Where $I_{tot} = I_+ + I_-$. Here the spin up (down) direction is always defined to be parallel (antiparallel) to the positive momentum direction. The spin polarization is equal to the asymmetry between $I_\uparrow$ and $I_\downarrow$:

$$P = \frac{I_\uparrow - I_\downarrow}{I_\uparrow + I_\downarrow} \quad (4)$$

All spin spectra were fitted to determine the intensity, energy position, and width of Lorentzian peaks. Each spin spectrum consists of Lorentzian peaks, the Shirley background, and the Fermi-Dirac distribution function. These are convoluted with a Gaussian that has a width of 80 meV to account for the energy-resolution of the spin slit. The equation used for fitting is as follows:

$$I = (((\sum_i L_i) + B_{shirley}) \times f(E)) * g(E) \quad (5)$$

Where $L_i$ represents the Lorentzians, $B_{shirley}$ is the Shirley background calculated by $B_{shirley} = \int \sum_i L_i \, dE$, $f(E)$ is the Fermi-Dirac distribution at room temperature, $*$ denotes convolution, and $g(E)$ is the normalized Gaussian with a width of 80 meV. To ensure the fitting is physically rigorous, the spin up and spin down spectra were fitted simultaneously with several constraints applied. Namely, the center and width of each Lorentzian was set to be equal for spin up and down spectra, as each Lorentzian represents the same band in both spectra; the width of Lorentzians at lower energy is constrained to be larger than those at higher energy, due to band width broadening which is primarily caused by electron-electron interactions[37,46]; $E_F$ was set to be equal for the two spectra.

To quantify the spin deviation angle from the spin data, the following procedure was used: (i) For the Kramers-Weyl cone and Weyl cone, the Lorentzian peak positions and widths were obtained from the fitted spin up and down intensities. (ii) The spin polarizations were integrated according to peak position (center) and width (range) for both $S_{//}$ and $S_\perp$. (iii) The deviation angle is calculated using arctan (integral of $P_\perp$ / integral of $P_{//}$). The complete process of quantification at azimuth angle 45° is laid out as an example in section 12 of the SI (Figure S12 and Figure S13).

The error for each spin angle is estimated using the following equation:

$$\sigma = \sqrt{\sigma_{fit}^2 + \sigma_{sys}^2} \tag{6}$$

Where $\sigma_{fit}$ is the random error from fitting and $\sigma_{sys}$ is the systematic error of the spin angle. For $\sigma_{fit}$, the error of peak position and peak width is used to estimate the error of the integral using a Monte Carlo method[47], elucidated in section 12 of the SI. Then the random error of deviation angle is obtained from integral errors using error propagation. For $\sigma_{sys}$ of the spin measurement, assuming that there is no spin deviation at 90° (There is a twofold screw axis along $\varGamma$-$Y$ enforcing parallel SML.), the experimental spin deviations at 90° for the Kramers-Weyl and Weyl cones are used as systematic errors, which were calculated as 7° and 6°, respectively.

**DFT calculations**

DFT calculations were performed using the full-potential local-orbital (FPLO) method in fully relativistic mode, within the generalized gradient approximation (GGA) using the Perdew-Burke-Ernzerhof (PBE) parametrization for the exchange-correlation functional[48,49]. A 12×12×12 $k$-point mesh was employed, and self-consistent convergence criteria of $10^{-6}$ eV per unit cell for both energy and charge density were achieved. Maximally localized Wannier functions were constructed from the DFT results, using valence orbitals for Rh (5s,6s,4d,5d,5p) and Si (3s,4s,3p,4p,3d). The topological properties were subsequently analyzed using the WannierTools package[50].

## Supplementary information

Additional information includes: DFT calculated type I Weyl point position and band structure along $\Gamma$-$X$; $\mathbf{k} \cdot \mathbf{p}$ model and Edelstein effect calculations; Comparison of spin textures at $E_F$ and $E - E_F$ = -0.2 eV; Additional soft X-ray ARPES data; $y$ and $z$ components of spin EDC in Figure 2; Time reversal symmetry of Weyl quasiparticle states; Matrix element effects along $\Gamma$-$Y$ at 20 eV and 40 eV; Out-of-plane spin component of Weyl cones at different azimuthal angles; $k_z$ scan for the cut at azimuthal angle 90°; Spin measurements at different azimuthal angles using 40 eV; Demonstration of spin deviation flip using the $\mathbf{k} \cdot \mathbf{p}$ model; Step-by-step process for extracting the spin angle at azimuthal angle 45°.

## Acknowledgements

J. W., A. K., M. P.-A., and D. G. acknowledge proposal Photons-251-13157 at the spin-ARPES end station of U125-PGM beamline at BESSY-II in Helmholtz Zentrum Berlin, Germany. J. W., M. P.-A. and A. K. acknowledge proposal I-20250544 at the SX-ARPES end station at the Beamline P04 of the PETRA III storage ring at DESY, Germany. E. C. M., and M. T. acknowledge proposal I-20230581 at the ASPHERE III end station at the Beamline P04 of the PETRA III storage ring at DESY, Germany. N.B.M.S. was funded by the European Union (ERC Starting Grant ChiralTopMat, project number 101117424). N.B.M.S acknowledges support by the German Research Foundation (DFG) as part of the German Excellence Strategy - EXC3112/1 - 533767171 (Center for Chiral Electronics) and Project-ID 328545488 of TRR 227. C. S. was financially supported by the European Union through EXQIRAL (No. 101131579), the Deutsche Forschungsgemeinschaft (DFG, German Research Foundation) through SFB 1143 (project ID 24731007), QUAST (project ID FOR 5249), and the Würzburg-Dresden Cluster of Excellence ctd.qmat – Complexity, Topology and Dynamics in Quantum Matter (EXC 2147, project ID 390858490). J. S.-B. acknowledges support from the Spanish AEI PID2024-157112OB-C53 (HYBRIDOS: HYPERFAN) and from the Comunidad de Madrid through projects TEC-2024/TEC-380 (Mag4TIC-CM). M. G. V. acknowledges the support of PID2022-142008NB-I00 funded by MICIU/AEI/10.13039/50110001 and FEDER, UE, the Canada Excellence Research Chairs Program for Topological Quantum Matter and to Diputacion Foral de Gipuzkoa Programa Mujeres y Ciencia. This work was also funded by the European Union NextGenerationEU/PRTR-C17.I1, as well as by the IKUR Strategy under the collaboration agreement between Ikerbasque Foundation and DIPC on behalf of the Department of Education of the Basque Government.

48. Perdew, J. P., Burke, K. & Ernzerhof, M. Generalized gradient approximation made simple. *Phys. Rev. Lett.* **77**, 3865–3868 (1996).

49. Koepernik, K. & Eschrig, H. Full-potential nonorthogonal local-orbital minimum-basis band-structure scheme. *Phys. Rev. B* **59**, 1743–1757 (1999).

50. Wu, Q., Zhang, S., Song, H.-F., Troyer, M. & Soluyanov, A. A. WannierTools: An open-source software package for novel topological materials. *Comput. Phys. Commun.* **224**, 405–416 (2018).

# Supplementary Information

# Quantifying quasiparticle chirality in a chiral topological semimetal

Jiaju Wang[1], Jaime Sánchez-Barriga[2,3], Amit Kumar[1], Markel Pardo-Almanza[1], Jorge Cardenas-Gamboa[4], Iñigo Robredo[5], Chandra Shekhar[6], Daiyu Geng[1], Emily C. McFarlane[1], Martin Trautmann[1,7], Enrico Della Valle[8], Moritz Hoesch[9], Meng-Jie Huang[9], Jens Buck[9], Vladimir N. Strocov[8], Annika Johansson[1,7,13], Stuart S. P. Parkin[1], Claudia Felser[6], Maia G. Vergniory[10,11,12] & Niels B. M. Schröter[1,7,13,*]

[1] Max Planck Institute of Microstructure Physics, Weinberg 2, 06120 Halle (Saale), Germany

[2] Helmholtz-Zentrum Berlin für Materialien und Energie, Elektronenspeicherring BESSY II, Albert-Einstein-Strasse 15, 12489 Berlin, Germany

[3] IMDEA Nanoscience, C/ Faraday 9, Campus de Cantoblanco, 28049 Madrid, Spain

[4] Leibniz Institute for Solid State and Materials Research, IFW Dresden, Helmholtzstraße 20, 01069 Dresden, Germany

[5] Smart Materials Unit, Luxembourg Institute of Science and Technology (LIST), Avenue des Hauts-Fourneaux 5, L-4362 Esch/Alzette, Luxembourg

[6] Max Planck Institute for Chemical Physics of Solids, 01187 Dresden, Germany

[7] Institute of Physics, Martin Luther University Halle-Wittenberg, 06120 Halle (Saale), Germany

[8] Swiss Light Source, Paul Scherrer Institut, Villigen, CH-5232 PSI, Switzerland

[9] Deutsches Elektronen-Synchrotron DESY, 22607 Hamburg, Germany

[10] Donostia International Physics Center, Manuel Lardizabal Ibilbidea 4, 20018 Donostia-San Sebastian, Gipuzkoa, Spain

[11] Département de Physique et Institut Quantique, Université de Sherbrooke, J1K 2R1 Sherbrooke, Québec, Canada

[12] Regroupement Québécois sur les Matériaux de Pointe (RQMP), H3T 3J7 Québec, Canada

[13] Halle-Berlin-Regensburg Cluster of Excellence CCE, 06120, Halle (Saale), Germany

*Corresponding author. Email address: niels.schroeter@mpi-halle.mpg.de

# Contents

# 1. DFT calculated type I Weyl point position and band structure along Γ-X

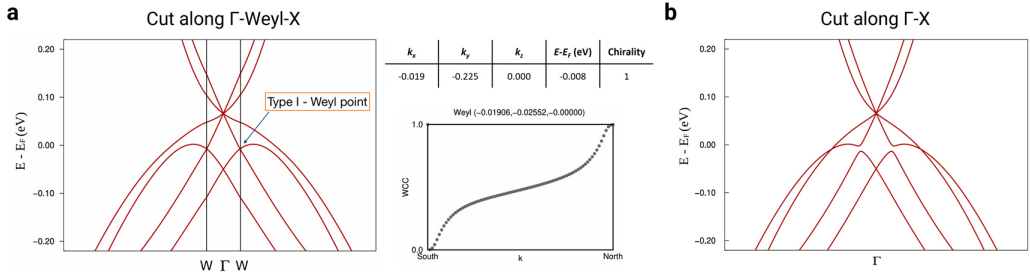

**Fig. S1: DFT calculations of the type I Weyl point and band structure along Γ-X. a**, Left panel shows band structure along the *k*-path Γ-type I Weyl point-X. Right panel shows the *k*-position, energy and chirality of the Weyl point and the Wilson loop plot of Wannier charge centers (WCC) vs. *k* loop. **b**, DFT calculated band structure along Γ-X.

Figure S1a shows the *k*-position and the energy of the type I Weyl point with chirality 1 calculated using DFT. Note that this topological chirality or chiral charge is defined by the flux of Berry curvature through a surface enclosing the node[1], and should be distinguished from the definition of electron chirality in the main text. The Wilson loop calculation shows a plot of Wannier charge centers *vs*. a closed parametrized *k*-loop. The Wannier charge centers increase monotonically from 0 to 1, indicating a Chern number of 1. Figure S1b shows the DFT-calculated band structure along Γ-X, where the main difference is that the band crossing for the type I Weyl point is not visible along this direction. Nevertheless, as can be seen from the contour plot in Figure 1b of main text, at $E - E_F = -0.2$ eV, we can probe the type I Weyl cone along multiple *k* directions.

# 2. $k \cdot p$ model and Edelstein effect calculations

To show that spin deviations can affect the magnitude of the Edelstein effect, a $k \cdot p$ method with SOC was adopted to model the band structure near Γ. A Dresselhaus-like anisotropy term was added to induce spin deviations. The full expression of the $k \cdot p$ Hamiltonian at Γ is:

$$H = \boldsymbol{k} \cdot \boldsymbol{L} + \lambda \, \boldsymbol{L} \cdot \boldsymbol{S} + \alpha \, \boldsymbol{f} \cdot \boldsymbol{L}$$

Where **L** represents angular momentum of a spin-1 multifold fermion, with

$$L_x = \begin{pmatrix} 0 & 0 & 0 \\ 0 & 0 & -i \\ 0 & i & 0 \end{pmatrix}, \quad L_y = \begin{pmatrix} 0 & 0 & i \\ 0 & 0 & 0 \\ -i & 0 & 0 \end{pmatrix}, \quad L_z = \begin{pmatrix} 0 & -i & 0 \\ i & 0 & 0 \\ 0 & 0 & 0 \end{pmatrix}.$$

$\lambda \boldsymbol{L} \cdot \boldsymbol{S}$ is the SOC term, where $\boldsymbol{S}$ is the physical spin operator. A similar Hamiltonian with SOC was used in the work by Tang et al[2]. $\alpha \boldsymbol{f} \cdot \boldsymbol{L}$ is the Dresselhaus-like anisotropy term. $\alpha$ (with unit eV·Å$^3$) determines the anisotropy strength, and $\boldsymbol{f}$ is given by[3]:

$$\boldsymbol{f}(\boldsymbol{k}) = \begin{pmatrix} k_x(k_y^2 - k_z^2) \\ k_y(k_z^2 - k_x^2) \\ k_z(k_x^2 - k_y^2) \end{pmatrix}$$

Note that the Dresselhaus-like anisotropy term is not an SOC term, but is a higher-order $k$ term in the Hamiltonian that has a form similar to the original Dresselhaus Hamiltonian. To prove that the Dresselhaus-like term $\boldsymbol{f} \cdot \boldsymbol{L}$ is allowed in the $\boldsymbol{k} \cdot \boldsymbol{p}$ Hamiltonian at $\Gamma$ in RhSi, it is sufficient to show that $\boldsymbol{f}$ transforms as a vector operator under the spatial symmetries of the point group (PG) at $\Gamma$, which is PG 23 (T), and that $\boldsymbol{f} \cdot \boldsymbol{L}$ obeys time reversal symmetry. The spatial symmetries of PG 23 are threefold rotation axes along $\langle 111 \rangle$ and twofold rotation axes along $\langle 100 \rangle$. Considering the threefold rotation along [111],

$$\boldsymbol{R}_{C3}\boldsymbol{f}(\boldsymbol{k}) = \begin{pmatrix} 0 & 0 & 1 \\ 1 & 0 & 0 \\ 0 & 1 & 0 \end{pmatrix} \begin{pmatrix} k_x(k_y^2 - k_z^2) \\ k_y(k_z^2 - k_x^2) \\ k_z(k_x^2 - k_y^2) \end{pmatrix} = \begin{pmatrix} k_z(k_x^2 - k_y^2) \\ k_x(k_y^2 - k_z^2) \\ k_y(k_z^2 - k_x^2) \end{pmatrix} = \boldsymbol{f}(\boldsymbol{R}_{C3}\boldsymbol{k})$$

For twofold rotation along [100],

$$\boldsymbol{R}_x\boldsymbol{f}(\boldsymbol{k}) = \begin{pmatrix} 1 & 0 & 0 \\ 0 & -1 & 0 \\ 0 & 0 & -1 \end{pmatrix} \begin{pmatrix} k_x(k_y^2 - k_z^2) \\ k_y(k_z^2 - k_x^2) \\ k_z(k_x^2 - k_y^2) \end{pmatrix} = \begin{pmatrix} k_x(k_y^2 - k_z^2) \\ k_y(k_x^2 - k_z^2) \\ k_z(k_y^2 - k_x^2) \end{pmatrix} = \boldsymbol{f}(\boldsymbol{R}_x\boldsymbol{k})$$

Similarly, it can be shown that $\boldsymbol{f}$ transforms as a vector under all symmetries of PG 23. The time reversal symmetry operator $\Theta$ sends $k$ to -$k$ and $L$ to -$L$. Therefore, we have:

$$\Theta [\boldsymbol{f}(\boldsymbol{k}) \cdot \boldsymbol{L}] \Theta^\dagger = [-\boldsymbol{f}(\boldsymbol{k})] \cdot [-\boldsymbol{L}] = \boldsymbol{f}(\boldsymbol{k}) \cdot \boldsymbol{L}$$

From the above proof, we conclude that the Dresselhaus-like anisotropy is allowed in our $\boldsymbol{k} \cdot \boldsymbol{p}$ Hamiltonian.

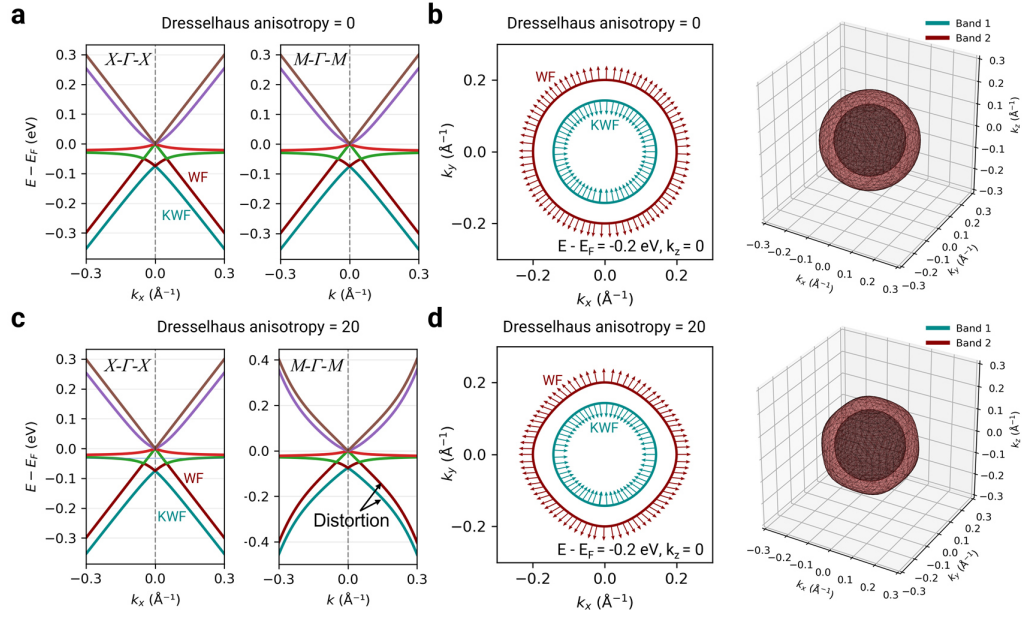

**Fig. S2: $k \cdot p$ model with tunable Dresselhaus anisotropy. a**, Band structure including SOC ($\lambda = 0.05$ eV) calculated using our $k \cdot p$ model with no anisotropy. **b**, Contour at $E - E_F = -0.2$ eV cutting through Kramers-Weyl and Weyl cones, with spin texture showing parallel SML. The 3D contour shows a spherical shape. **c**, Band structure with Dresselhaus anisotropy, $\alpha$ was set to be 20. Distortion in the band structure is visible along $\Gamma$-$M$. **d**, Contour at $E - E_F = -0.2$ eV with $\alpha = 20$ eV·Å$^3$ now shows distortions from the circular shape, and spin deviations from parallel SML which are comparable to that in DFT calculations. The 3D contour now has a distorted spherical shape.

Using this model, the calculated band structures and contours together with the spin texture are shown in Figure S2. The SOC parameter $\lambda$ was set to be 0.05 eV. The spin of band $n$ is calculated using the spin expectation value:

$$s(k) = \langle \psi_n(k) | S | \psi_n(k) \rangle$$

It can be seen that when the Dresselhaus anisotropy coefficient $\alpha = 20$ eV·Å$^3$, at $E - E_F = -0.2$ eV, the contour shapes and spin deviations of the Kramers-Weyl and Weyl cones are comparable to those calculated with DFT in the main text.

The NECD was integrated over the band contour surface to obtain a normalized electron chirality, which was labelled as integrated |NECD| in Figure 1 of the main text. This value was found to decrease with increasing $\alpha$. The Edelstein response at $E - E_F = -0.2$ eV was calculated using the equation for spin density[4]:

$$s_{Edel} = \frac{e}{V^d} \sum_k \frac{\partial f_k^0}{\partial \varepsilon} \tau_k s_k (v_k \cdot E)$$

Which was converted into an integral over the contour surface at zero temperature. Assuming a constant momentum relaxation time $\tau_k$, we calculated a normalized Edelstein response with respect to $\alpha = 0$, setting the magnitude of $s_{Edel}$ as 1 for $\alpha = 0$, and calculating the relative magnitudes at different $\alpha$ values. It was found that the response decreased with increasing $\alpha$. Since the NECD decreases with $\alpha$, we conclude that the Edelstein response decreases with decreasing NECD in the main text. Due to the symmetries in RhSi, the Edelstein response should be homogeneous. This homogeneity was tested by changing the direction of the electric field $E$ from $x$ to $y$, $z$, and a random direction. The results showed that the magnitude of the Edelstein response remained the same, further corroborating that our Hamiltonian complies with the symmetries of PG 23.

## 3. Comparison of spin textures at $E_F$ and $E - E_F = -0.2$ eV

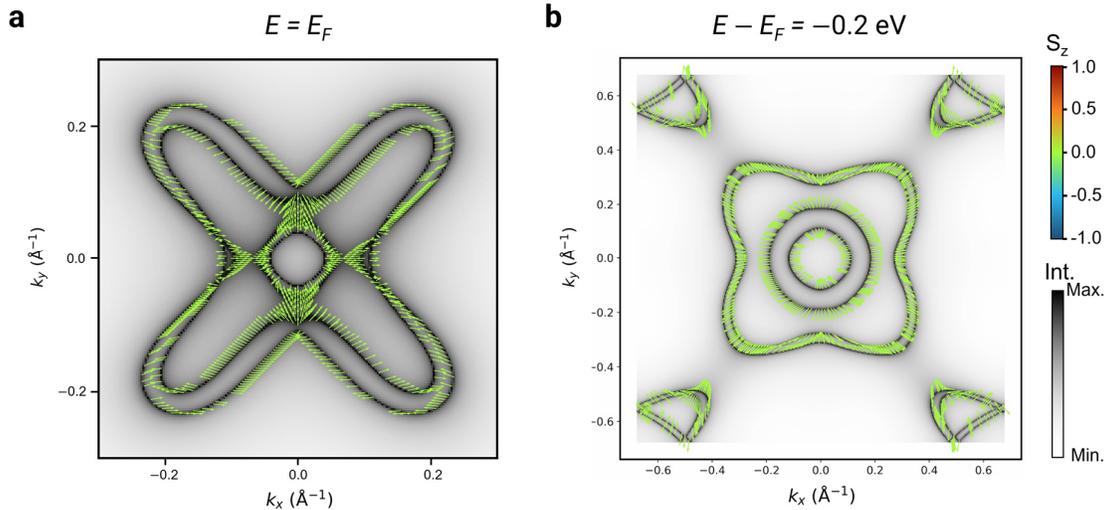

**Fig. S3 Spin textures at $E_F$ and $E - E_F = -0.2$ eV.** Constant energy contours with spin textures at **a,** $E_F$ and **b,** $E - E_F = -0.2$ eV. Comparing these two spin textures, it can be seen that the spin deviation magnitude at $E_F$ is similar to that at $E - E_F = -0.2$ eV. Thus, we expect the spin deviations at $E_F$ to cause a decrease in Edelstein response similar to that calculated at $E - E_F = -0.2$ eV.

## 4. Additional soft X-ray ARPES data

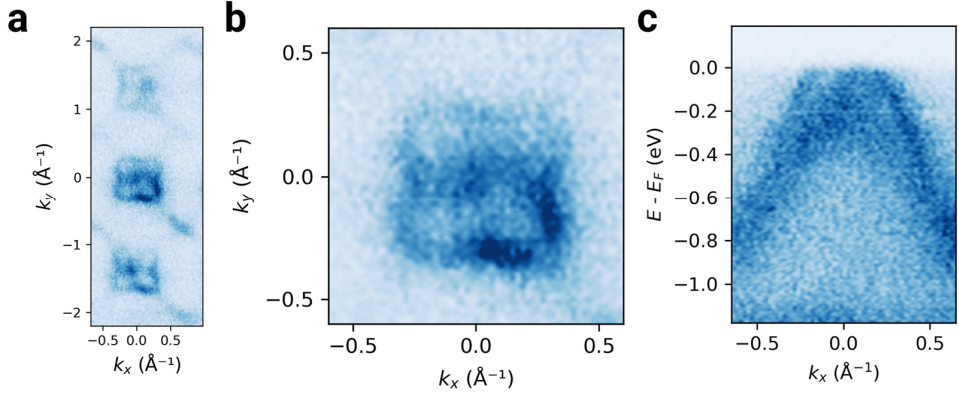

**Fig. S4: Soft X-ray ARPES data measured at 540 eV. a,** FS over several Brillouin zones taken at 540 eV. **b,** FS of the first Brillouin zone. **c,** cut along $\Gamma$-$X$.

## 5. $y$ and $z$ components of spin EDC in Figure 2

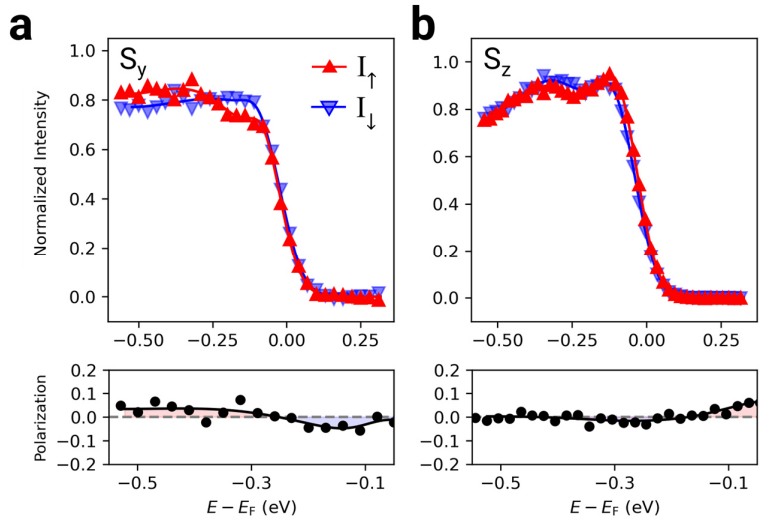

**Fig. S5: $y$ and $z$ components of spin EDC taken from cut along $\Gamma$-$X$. a,** $y$ component spin data corresponding to the EDC taken from the cut along $\Gamma$-$X$ in Figure 2 of the main text. **b,** $z$ component spin data for the same EDC.

From Figure S5, we can see that the splittings of Kramers-Weyl and Weyl cones are not visible in $y$ and $z$ spin components, and the polarizations are smaller compared to that of the $x$ direction.

# 6. Time reversal symmetry of Weyl quasiparticle states

To test the time reversal symmetry of the states on Kramers-Weyl and Weyl cones, we probe the spin at two opposite $k$ points along the $\Gamma$-$Y$ direction of the VUV bulk state at 40 eV, which avoids the Fermi arc surface states. Spin EDCs i and ii are both approximately 0.1 Å$^{-1}$ away from $\Gamma$ along $\Gamma$-$Y$. Their positions are marked with brown and blue dashed lines on the cut along $\Gamma$-$Y$ in Figure S6a. By measuring the reversal of the bulk state spin direction for the two EDCs, we can verify the time reversal symmetry in RhSi, and confirm the SOC-origin of the measured spin signal.

Figure S6b shows the spin results for EDC i. The spin spectra only show one SOC-split band, corresponding to the KW+W band in the main text. After fitting, a splitting of 71 ± 4 meV is extracted from $S_y$. The Kramers-Weyl cone has a peak around 0.18 eV in the spin down intensity of $S_y$, and the Weyl cone has a peak around 0.11 eV in the spin up intensity. This is consistent with the spin texture obtained from DFT calculations. For $S_x$, though the polarization is smaller than $S_y$, spin splitting is visible in the spin up and down intensities. This could be due to a systematic error such that EDC i is not strictly on the $\Gamma$-$Y$ line, leading to deviations in the $x$ direction. For $S_z$, the spin polarization is nearly zero, and there is no clear splitting between spin up and down intensities.

Figure S6c shows the results for EDC ii. For $S_y$, a splitting of 77 ± 9 meV is calculated. The $y$ components of spin for the Kramers-Weyl and Weyl cones are spin up and spin down, respectively. While in EDC i, it is spin down and spin up. In the $x$ component, the polarization detected for EDC ii also arises from a systematic error. Comparing to the $x$ component of EDC i, the spin is flipped in a fashion similar to that in $S_y$. The $S_z$ component shows almost no spin polarization.

Overall, the spin EDCs i and ii show that the spin direction of the Kramers-Weyl cone and the Weyl cone is flipped at opposite $k$ points, confirming that time reversal symmetry is preserved, and the origin of the spin measured is indeed SOC, rather than final state effects or residual magnetism.

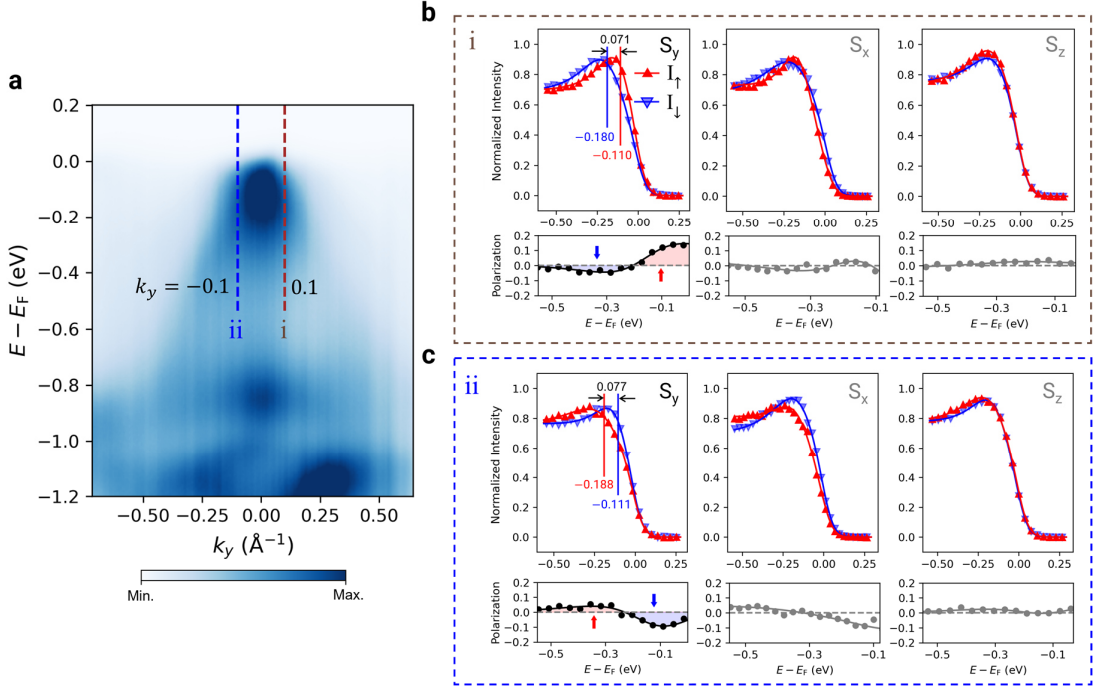

**Fig. S6: Time reversal symmetry of Weyl quasiparticle states in RhSi. a**, Cut along $\Gamma$-$Y$ taken at 40 eV. Spin EDCs i and ii were taken at opposite $k$ points to test time reversal symmetry. **b**, Spin data for EDC i, highlighting the spin in $k_y$ direction. The splitting is 71 ± 4 meV. **c**, Spin data for EDC ii, highlighting the spin in $k_y$ direction. A splitting of 77 ± 9 meV is extracted. The splitting values are consistent with each other and that in Figure 2 within the errors. In the $y$ direction, the spin reverses for spin EDC i and ii, indicating that time reversal symmetry is preserved.

## 7. Matrix element effects along $\Gamma$-$Y$ at 20 eV and 40 eV

Figure S7a shows that at 20 eV, the matrix element effect causes the bands at positive $k_y$ values to be barely visible. Nevertheless, at 20 eV the Kramers-Weyl and Weyl cones show high intensity (Fig. S7b), which is advantageous for spin-resolved measurements. At 40 eV, there is high intensity near $E_F$ close to the Kramers-Weyl point, and low intensity in the Kramers-Weyl and Weyl cones. Therefore, 20 eV was chosen for the spin measurements at different azimuthal angles.

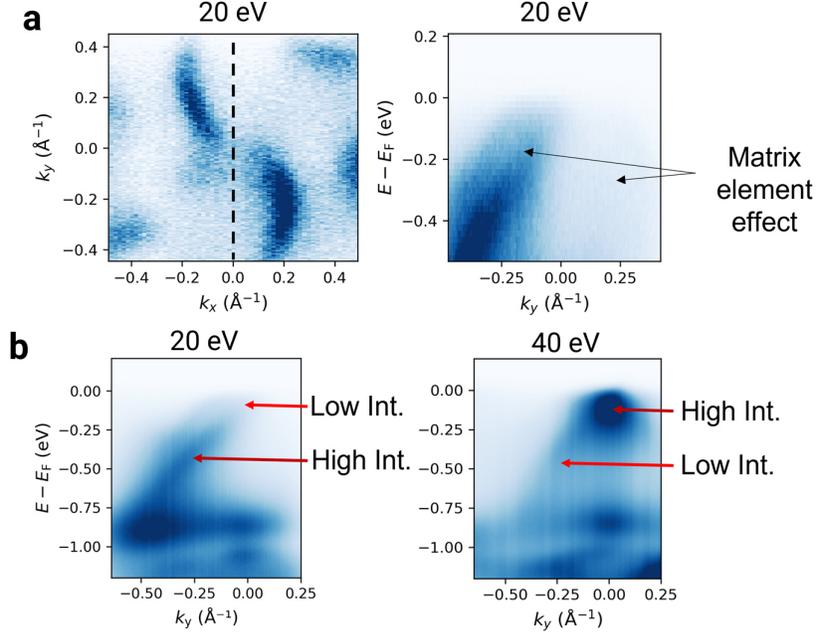

**Fig. S7: Matrix element effects along *Γ-Y* at 20 eV and 40 eV. a,** Left panel shows FS at 20 eV. Along *Γ-Y*, the intensity at negative $k_y$ values is higher than that at positive $k_y$. Right panel shows the cut through *Γ-Y*. Due to the matrix element effect, the bands at positive $k_y$ values are not visible. **b,** Comparison of cuts through *Γ-Y* at 20 eV and 40 eV.

## 8. Out-of-plane spin component of Weyl cones at different azimuthal angles

Figure S8 shows the out-of-plane spin component of the RhSi bulk band at different azimuthal angles. From Figure S8, it can be seen that apart from 45°, the splitting between spin up and down intensities is small, and the magnitude of the polarization is less than 5%. This small $z$ component is found to have negligible effect on the spin deviation angles calculated in the main text. For 45°, although the out-of-plane component is larger than those at other angles, the polarization is still small compared to that of in-plane spin components shown in Figure 3. Therefore, the in-plane spin spectra are used to calculate the spin angles in the main text.

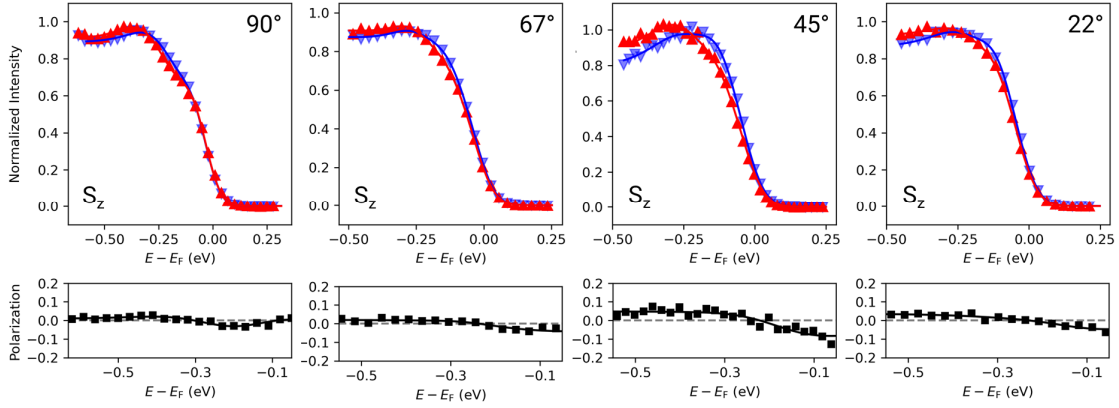

**Fig. S8: Out-of-plane spin of the RhSi bulk state near $\Gamma$ at different azimuthal angles.** Top row shows spin up and down intensities from 90° to 22° and the bottom row shows the corresponding spin polarizations.

## 9. $k_z$ scan for the cut at azimuthal angle 90°

Figure S9a shows the cut at azimuthal angle 90° for RhSi (001) taken at 20 eV. It can be seen that in the cut there is one surface state (labelled with S) at higher energy, separated from the KW+W band. In Figure S9b, the surface state remains the same with increasing $k_z$, corroborating its surface nature. The KW+W band changes with increasing $k_z$, indicating that it is a bulk state.

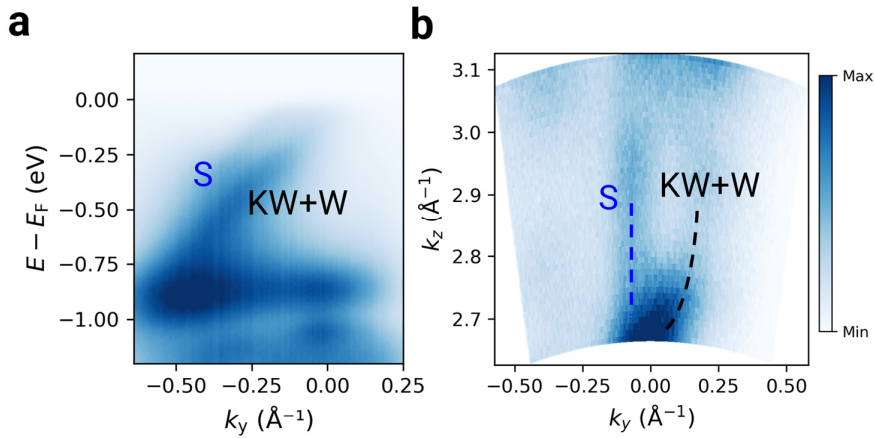

**Fig. S9: $k_z$ scan along $\Gamma$-$Y$ showing separated surface and bulk states. a**, Cut along $\Gamma$-$Y$ at 20 eV, showing that the surface and bulk states are differentiable. **b**, $k_z$ scan along $\Gamma$-$Y$. The surface state shows no change with $k_z$, while the bulk state changes, corroborating the surface and bulk nature of the bands.

## 10. Spin measurements at different azimuthal angles using 40 eV

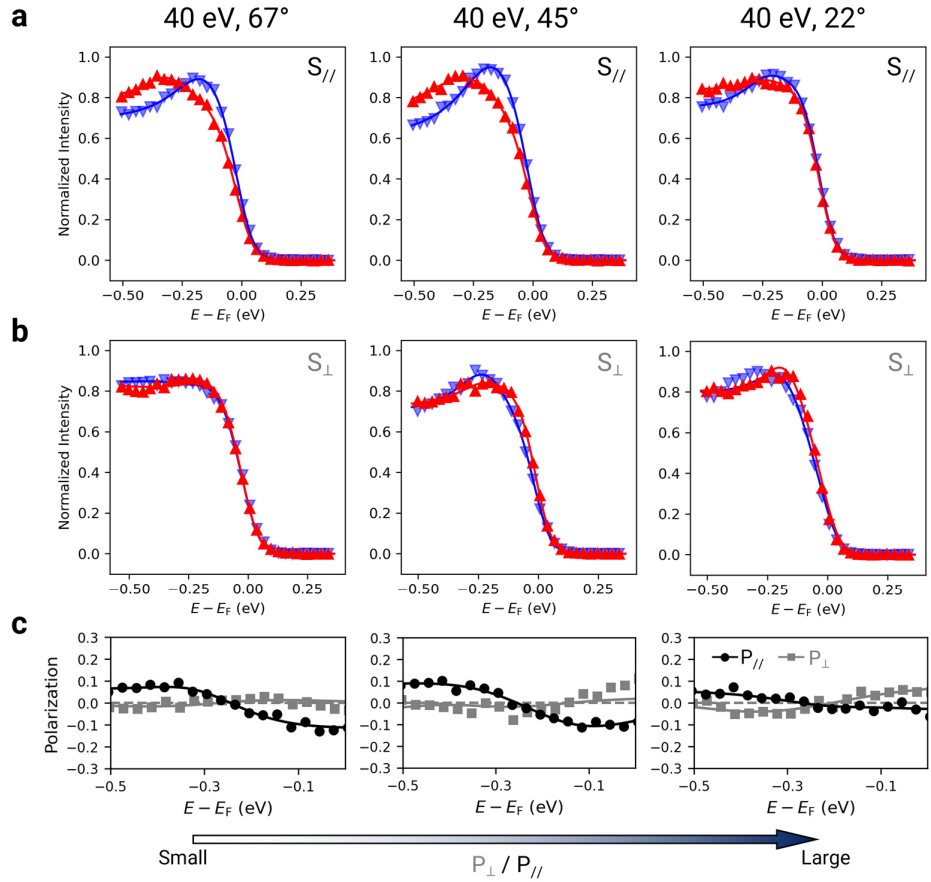

**Fig. S10: Spin EDC measured at azimuthal angles 67° to 22° using 40 eV photons. a,** Parallel-to-momentum component of in-plane spin up and down intensities at different azimuth angles. **b,** Perpendicular-to-momentum component of in-plane spin up and down intensities. **c,** Spin polarizations of both in-plane components. The ratio $P_\perp/P_{//}$ increases from 67° to 22°, indicating an increasing spin deviation.

Figure S10 shows spin EDCs measured using 40 eV photons at the same azimuthal angles (67°, 45°, and 22°) and at $k$ values similar to those in Figure 3. The shapes and asymmetries of spin intensities in Figure S10 a and b are comparable to those in Figure 3. In Figure S10c, the spin polarizations have the same sign with magnitudes similar to those in Figure 3. The ratio $P_\perp/P_{//}$ increases with decreasing azimuthal angle, indicating an increase in spin deviation from parallel SML, consistent with conclusions drawn from Figure 3. The above analysis suggests that at 20 eV and 40 eV, we are probing the initial-state spin texture of RhSi Weyl cones.

## 11. Demonstration of spin deviation flip using the $k \cdot p$ model

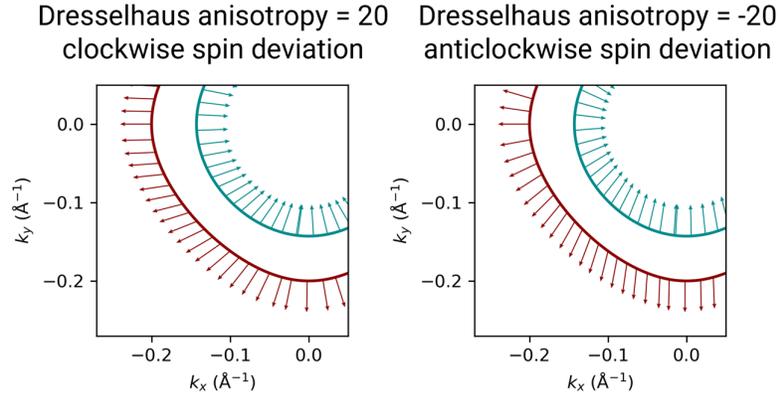

**Fig. S11: Spin deviation flips with Dresselhaus-like anisotropy sign change.** Spin textures obtained using the $k \cdot p$ model with Dresselhaus-like term coefficient 20 and -20. The spin deviation flips from clockwise to anticlockwise.

## 12. Step-by-step process for extracting the spin angle at azimuthal angle 45°

This section carefully lays out the calculation of spin angle at azimuthal angle 45° as an example to demonstrate the data analysis process.

### 12.1 Fitting using python package 'lmfit'

The spin up and spin down data were fitted together with Equation 5 in the main text using the python fitting package 'lmfit', which also computes errors for the fitting parameters. The simultaneous fitting ensures that the energy positions and widths are the same for the same Lorentzian in spin up and spin down spectra. Figure S12 demonstrates the fitting process, displaying different components of the fit.

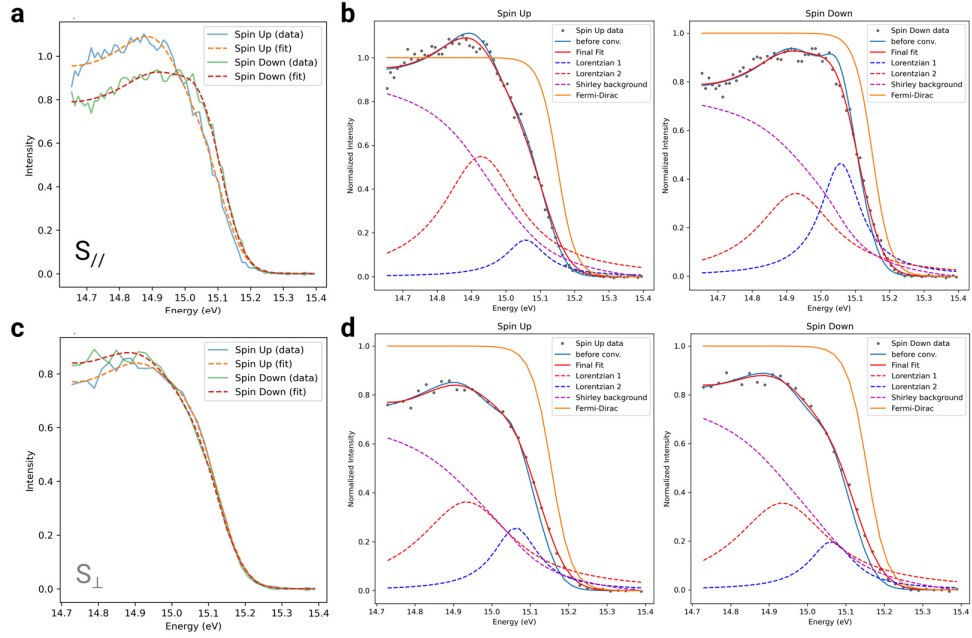

**Fig. S12: Fitting spin up and down spectra. a, c,** Spin up and down intensities in the direction **a,** parallel to momentum and **c,** perpendicular to momentum. **b, d,** Components of the fit for spin up and down spectra **b,** parallel and **d,** perpendicular to momentum.

**12.2 Polarization integration and error propagation using a Monte Carlo method**

A fitted polarization was obtained from fitted spin up and spin down intensities. The polarization was integrated according to the peak energy and width of the Lorentzians. To extract an error for the integrals, a Monte Carlo method was used, which calculates the integral 1000 times with the center and width randomly varying within their respective errors. Then the standard deviation of the results is used to obtain an error for the integral. This process is illustrated in Figure S13.

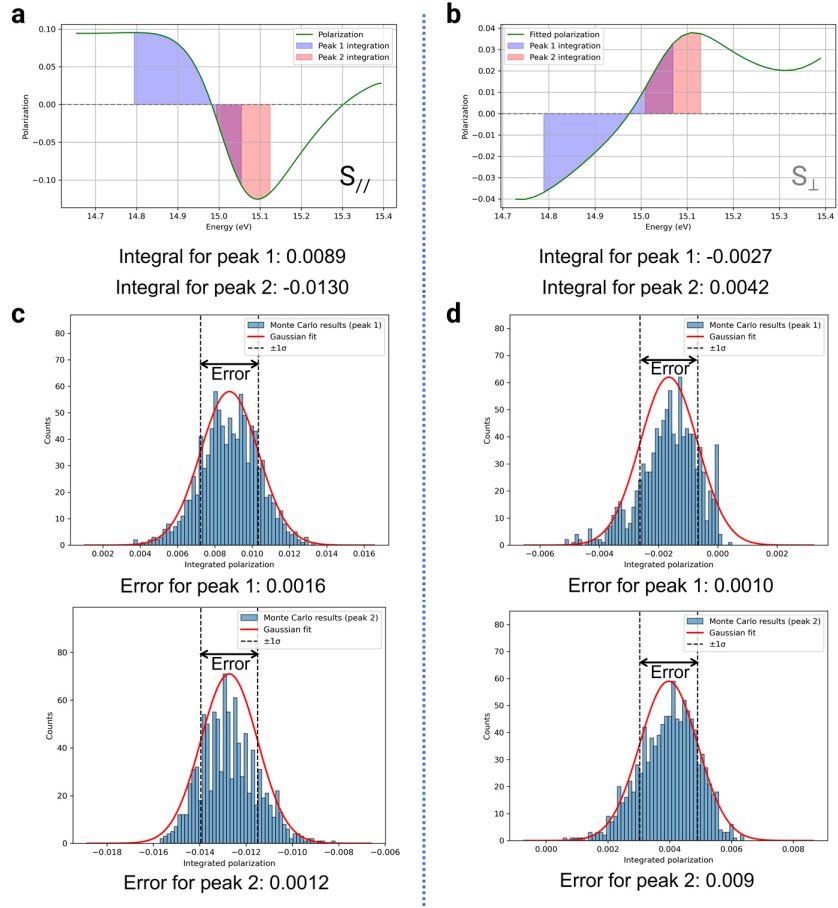

**Fig. S13: Integration of polarization and Monte Carlo error analysis. a, b,** Integration of fitted polarization based on Lorentzian peak energy and width for **a,** parallel, and **b,** perpendicular in-plane components. Histograms illustrating the Monte Carlo integration results for **c,** parallel, and **d,** perpendicular in-plane components. The error of the integral is estimated by the energy difference between $\pm \sigma$ of the Gaussian fitted to the Histogram.

## 12.3 Spin angle calculation

After the Monte Carlo analysis, we have obtained values with errors for integrals of the polarization. These integrals represent the relative magnitudes of spin up and spin down in parallel and perpendicular directions, and can be used to obtain an in-plane spin deviation angle using arctan(|Integral$_{//}$ / Integral$_\perp$|). The values were calculated to be:

$$\text{Peak 1: arctan}\left(\frac{0.0027 \pm 0.0010}{0.0089 \pm 0.0016}\right) = 17° \pm 7°$$

$$\text{Peak 2: arctan}\left(\frac{0.0042 \pm 0.0009}{0.0130 \pm 0.0012}\right) = 18° \pm 4°$$

Peak 1 is the Lorentzian at lower energy corresponding to the Kramers-Weyl cone, and peak 2 corresponds to the Weyl cone. Combining these results with the systematic error of 6° for the Kramers-Weyl cone and 7° for the Weyl cone, the spin deviation angle at azimuthal angle 45° for the Kramers-Weyl cone is 17° ± 9°, and that of the Weyl cone is 18° ± 8°, as summarized in Table 1 of the main text.

**Supplementary References**